# Magneto-Rayleigh-Taylor instability driven by a rotating magnetic field

**Shu-Chao Duan** [a,b,*] *et al*

[a] *Institute of Fluid Physics, CAEP, P. O. Box 919-105, Mianyang 621999, China*

[b] *Department of Modern Physics, University of Science and Technology of China, Hefei 230026, China*

**Corresponding author: s.duan@163.com*

套筒(liner) ，厚度 *h*，外半径为 *a*，内半径为 *a-h*。柱坐标系。套筒内外均有磁场 $B_i = (0, B_{\phi i}, B_{zi})$ 和 $B_o = (0, B_{\phi o}, B_{zo})$ 。在随体坐标系中，等效重力加速度为 $\mathbf{g_r} = g\mathbf{e_r}, g > 0$ 。力平衡条件

$$\rho g h = \frac{B_o^2 - B_i^2}{2} \equiv \frac{\Delta B^2}{2} \equiv p_m^+ - p_m^-. \tag{1}$$

不可压流体近似。线性化的运动方程

$$\rho \ddot{\xi}^{\pm} - \rho g \partial_r \xi^{\pm} + \partial_r \delta p_m^{\pm} = 0. \tag{2}$$

真空区扰动磁场可以表示为 $\delta B = \nabla \psi$ ，它满足拉普拉斯方程 $\nabla^2 \psi = 0$ 。对于 $\phi$ 和 *z* 因子是谐波的情形，其径向的解是修正 Bessel 函数。为了保持有限性，内区取第一类修正 Bessel 函数，外区取第二类修正 Bessel 函数。因此，内外区的扰动磁场为

$$\begin{aligned} &\delta B_{ri} = kC_i I_-', \delta B_{\phi i} = \tfrac{im}{a} C_i I_-, \delta B_{zi} = ikC_i I_-, \\ &\delta B_{ro} = kC_o K', \delta B_{\phi o} = \tfrac{im}{a} C_o K, \delta B_{zo} = ikC_o K, \end{aligned} \tag{3}$$

其中记号

$$I_- \equiv I_m(k(a-h)), I_-' \equiv I_m'(k(a-h)), K \equiv K_m(ka), K' \equiv K_m'(ka), \tag{4}$$

并且导数是对宗量的。根据磁场平行于扰动界面的条件可以得出

$$C_i = \frac{imB_{\phi i} + ik(a-h)B_{zi}}{k(a-h)I_-'} \xi^-, C_o = \frac{imB_{\phi o} + ikaB_{zo}}{kaK'} \xi^+. \tag{5}$$

扰动界面上的磁压为 $\delta p_m = B_\phi \delta B_\phi - B_\phi^2 \xi / a + B_z \delta B_z$ ，内外界面处分别为

$$\delta p_m^- = \rho g f_I \xi^-, \delta p_m^+ = \rho g f_K \xi^+, \tag{6}$$

其中记号

$$f_I \equiv -\frac{2hq^2 B_i^2 I_- \cos^2\theta_i}{k\Delta B^2 I_-^{'}} - \frac{2hB_{\phi i}^2}{(a-h)\Delta B^2}, f_K \equiv -\frac{2hq^2 B_o^2 K \cos^2\theta_o}{k\Delta B^2 K^{'}} - \frac{2hB_{\phi o}^2}{a\Delta B^2}, \tag{7}$$

其中 $q=\sqrt{(m/a)^2+k^2}$ ， $\theta=\theta(t)$ 是 $q$ 与 $B$ 之间的夹角。

代入方程(2)之后得到

$$\ddot{\xi}^- - g(1-f_I)\partial_r \xi^- = 0, \tag{8}$$

$$\ddot{\xi}^+ - g(1-f_K)\partial_r \xi^+ = 0. \tag{9}$$

由于 $\nabla\cdot\xi=\nabla\times\xi=0$ ，所以其通解是

$$\xi = [AI_m(ka)+BK_m(ka)]e^{im\phi+ikz}\xi(t). \tag{10}$$

代入(8)(9)式，得到关于 A、B 的代数方程。要求相应的行列式为零，得出

$$\ddot{\xi}^2 + Rgk\ddot{\xi}\xi - Sg^2k^2\xi^2 = 0, \tag{11}$$

其中

$$R = -\frac{(1-f_I)(I_-^{'}K - IK_-^{'}) + (1-f_K)(I_-K^{'} - I^{'}K_-)}{I_-K - IK_-}, \quad S = -\frac{(1-f_I)(1-f_K)(I_-^{'}K^{'} - I^{'}K_-^{'})}{I_-K - IK_-}. \tag{12}$$

前面(11)式即

$$(\ddot{\xi} - \gamma_+ gk\xi)(\ddot{\xi} - \gamma_- gk\xi) = 0, \tag{13}$$

其中

$$\gamma_\pm = -\tfrac{R}{2} \pm \sqrt{\tfrac{R^2}{4} + S}. \tag{14}$$

**WKB 的两种退化情况：**

**薄壳退化：**若 $kh\to 0$ 且 $h\to 0$ ，则

$$f_I \to -kh\frac{2q^2B_i^2 I\cos^2\theta_i}{k^2\Delta B^2 I^{'}} - kh\frac{2B_{\phi i}^2}{ka\Delta B^2} \to 0, \quad f_K = -kh\frac{2q^2B_o^2 K\cos^2\theta_o}{k^2\Delta B^2 K^{'}} - kh\frac{2B_{\phi o}^2}{ka\Delta B^2} \to 0, \tag{15}$$

$$
\begin{aligned}
R &= -\frac{(I_-'K - IK_-') + (I_-K' - I'K_-)}{I_-K - IK_-} + \frac{(I_-'K - IK_-')}{I_-K - IK_-} f_I + \frac{(I_-K' - I'K_-)}{I_-K - IK_-} f_K \\
&\rightarrow \frac{1}{ka} + \frac{2q^2 B_i^2 I \cos^2\theta_i}{k^2 \Delta B^2 I'} + \frac{2B_{\phi i}^2}{ka\Delta B^2} - \frac{2q^2 B_o^2 K \cos^2\theta_o}{k^2 \Delta B^2 K'} - \frac{2B_{\phi o}^2}{ka\Delta B^2} \\
&= \frac{2q^2 B_i^2 I \cos^2\theta_i}{k^2 \Delta B^2 I'} - \frac{2q^2 B_o^2 K \cos^2\theta_o}{k^2 \Delta B^2 K'} + \frac{B_{\phi i}^2 + B_{zo}^2 - B_{\phi o}^2 - B_{zi}^2}{ka\Delta B^2}, \\
S &\rightarrow -\frac{(I_-K' - I'K_-)}{I_-K - IK_-} \rightarrow \frac{q^2}{k^2}.
\end{aligned}
\tag{16}
$$

若内部无磁场，并取 WKB 近似，则(11)式与 Ryutov 的 WKB 结果 (36)式一致。

**平板退化**：若 $a \rightarrow \infty$，并取 WKB 近似，则 $R$ 和 $S$ 退化成 Lau 的(7)式。若进一步内部无磁场，则(11)式退化成 Harris 的(26)式，即

$$\Gamma_{1,2}^2 = \gamma_+ gk, \Gamma_{3,4}^2 = \gamma_- gk, \tag{17}$$

$$\gamma_\pm = -\tfrac{R}{2} \pm \sqrt{\tfrac{R^2}{4} + S}, \tag{18}$$

$$R = 2kh\coth(kh)\cos^2\theta, S = 1 - 2kh\cos^2\theta, \tag{19}$$

其中 $\theta$ 为旋转磁场 $B(t)$与波矢 $k=(kx,ky,0)$之间夹角 $\theta(t)=\omega_b t+\theta_0$。恒有 $R \geq 0, \frac{R^2}{4} + S \geq 0$，因此 $\gamma_-$ 恒负，是振荡模式。仅当 $S > 0$ 时，$\Gamma_{1,2}^2 > 0$，存在不稳定的正根，因此定义增长率 $\Gamma = \sqrt{\max(0, \Gamma_{1,2}^2)}$，及归一化增长率 $\gamma = \Gamma / \sqrt{gk} = \sqrt{\max(0, \gamma_+)}$ 。

若磁场方向的旋转频率 $\omega_b \ll \Gamma$，则在 $T \gg 2\pi / \omega_b$ 的时间尺度上，每个模式 $\vec{k} \equiv (k, \theta)$ 都感受到一个与 $\theta$ 无关的平均增长率 $\overline{\Gamma} = \langle \Gamma \rangle_\theta$ 及归一化平均增长率 $\overline{\gamma} = \langle \gamma \rangle_\theta$。下图依次是 $\gamma$ 和 $\overline{\gamma}$：

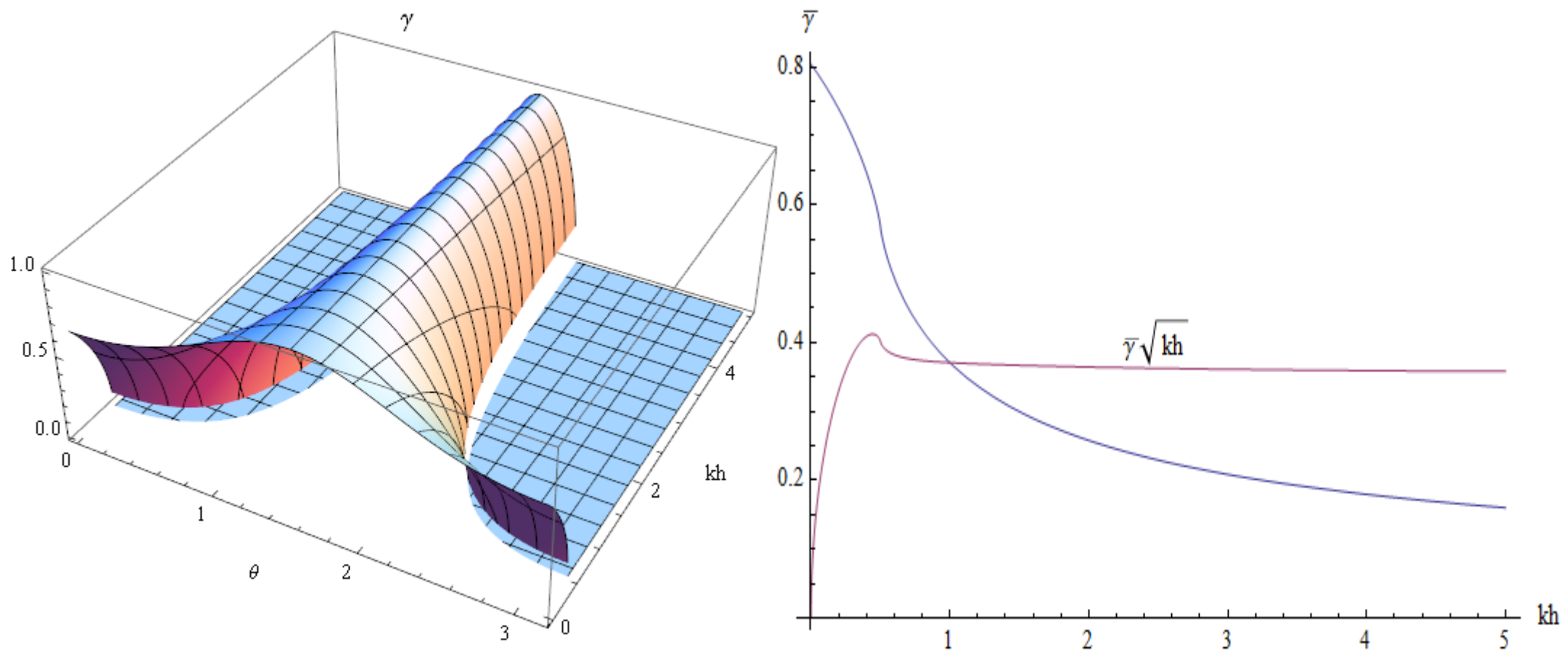

其右端极限：$kh \rightarrow \infty$ 时，$\overline{\gamma} \rightarrow \frac{\sqrt{2}}{4} / \sqrt{kh}$，即 $\overline{\Gamma} \rightarrow \frac{\sqrt{2}}{4}\sqrt{g/h}$ 。

其左端极限：$kh \to 0$ 时，$\bar{\gamma} \to 0.8041731875209983 < 1$，表明即使在最左端亦存在抑制作用。

极大值出现在：$kh = 0.44329236527950494$ 时，$\bar{\Gamma} = 0.4122332114675114\sqrt{g/h}$ 。

若 $\omega_b \gg \Gamma$，则在 $T \gg 2\pi/\omega_b$ 的时间尺度上，感受到的是平均值 $\bar{R} = \langle R\rangle_\theta, \bar{S} = \langle S\rangle_\theta$，将 $R$，$S$ 的表式中的 $\cos^2\theta$ 替换为 1/2 即得到 $\bar{R}, \bar{S}$ 的表式，进而得到 $\bar{\Gamma}$ 及 $\bar{\gamma}$ 的表式。

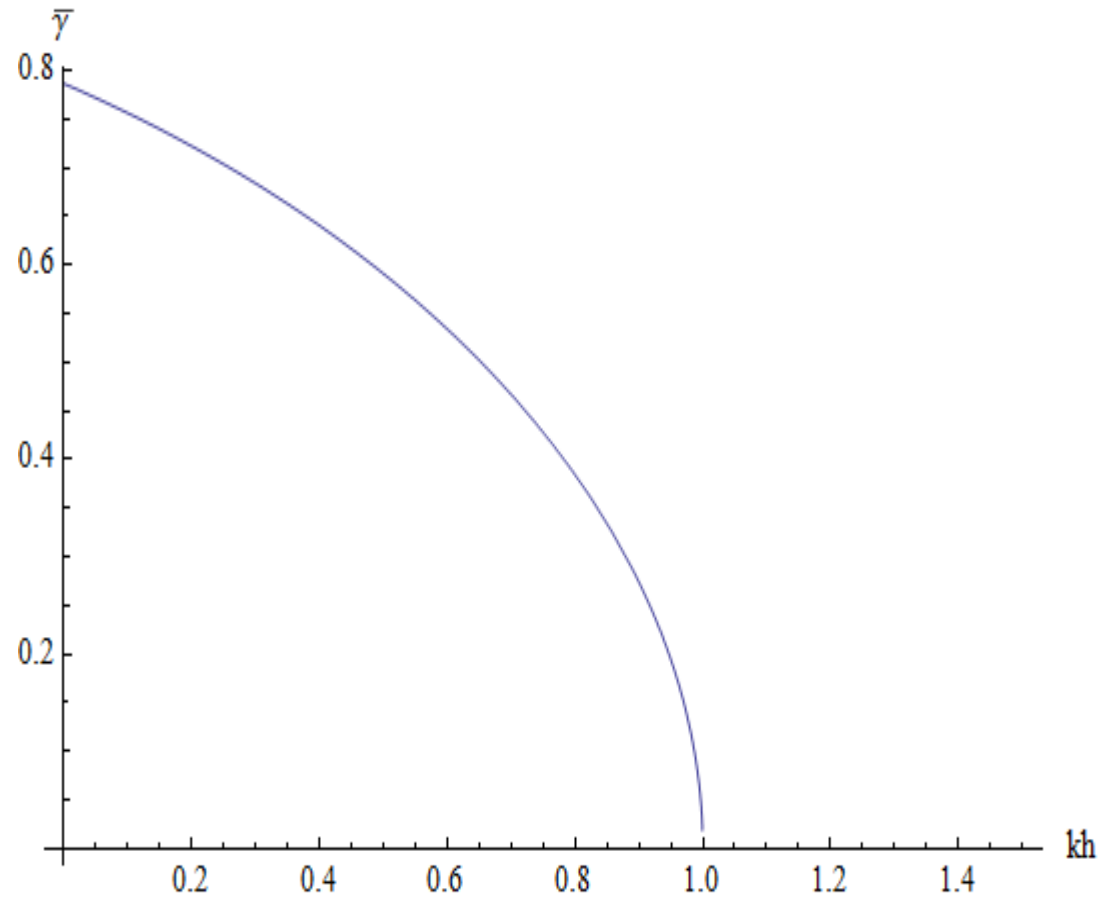

其左端极限是 $\sqrt{\frac{\sqrt{5}-1}{2}} = 0.7861513777574233$ 。$kh \geq 1$ 的模式完全致稳。

归一化增长率既代表了对经典解 $\sqrt{gk}$ 的相对值，因此可以看出，在 $T > 2\pi/\Gamma$ 的时间尺度上，磁场方向旋转对所有模式都有抑制作用，这可能表明，Θ 箍缩和 Z 箍缩交替进行的构型可能有助于抑制 MRT。归一化增长率又代表了对于在固定加速度 $g$ 驱动下的某一固定模式，例如 $g$=1,$k$=1，增长率随厚度的依赖关系，这时将横轴理解为绝对厚度 $h$，因此可以看出厚度具有单调的抑制作用，而且，只要足够厚，可以抑制到任意程度，没有饱和。若总质量固定，$\rho \downarrow$，则 $h\uparrow$，则增长率单调下降，表明低密度实心套筒有助于抑制 MRT。

这些结果能退化至半无限平面即无限厚的结果。$kh \to \infty$，则 $R \to 2kh\cos^2\theta, S = 1 - 2kh\cos^2\theta$，因此 $\Gamma_{1,2}^2 = gk - 2ghk^2\cos^2\theta = gk - V_A^2 k^2 \cos^2\theta$：

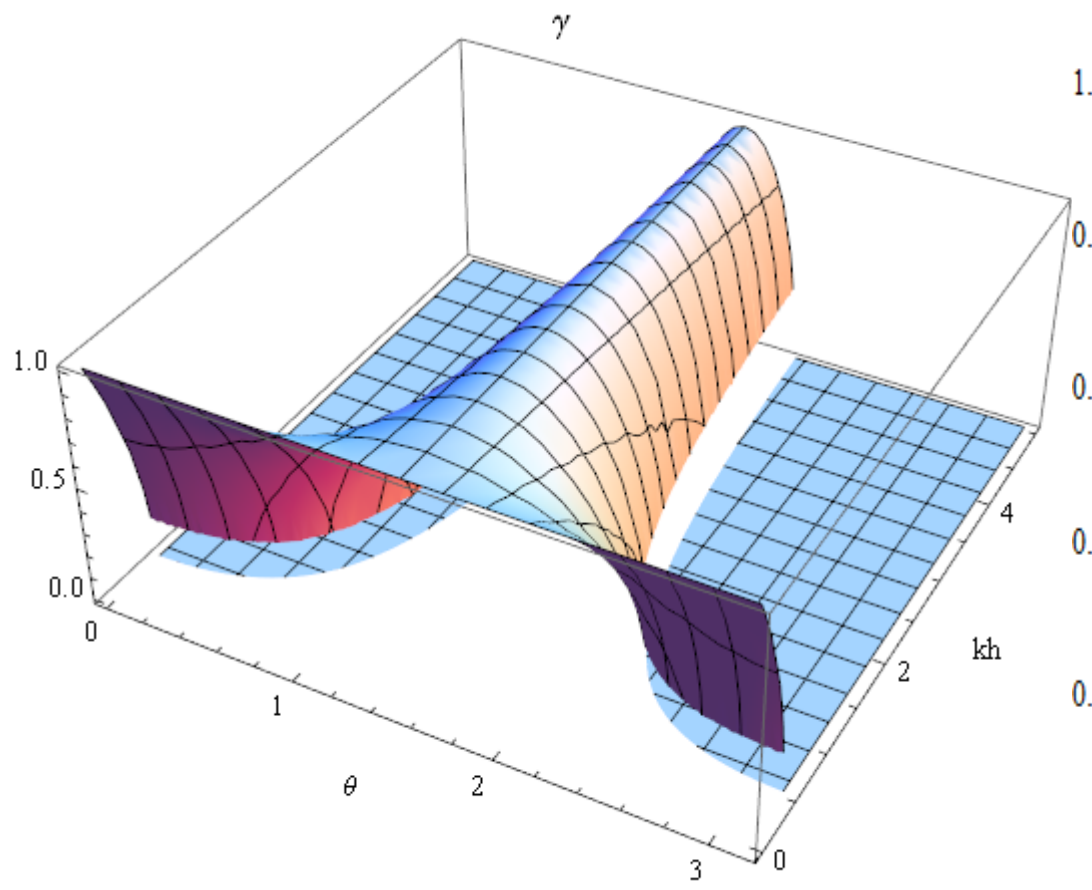

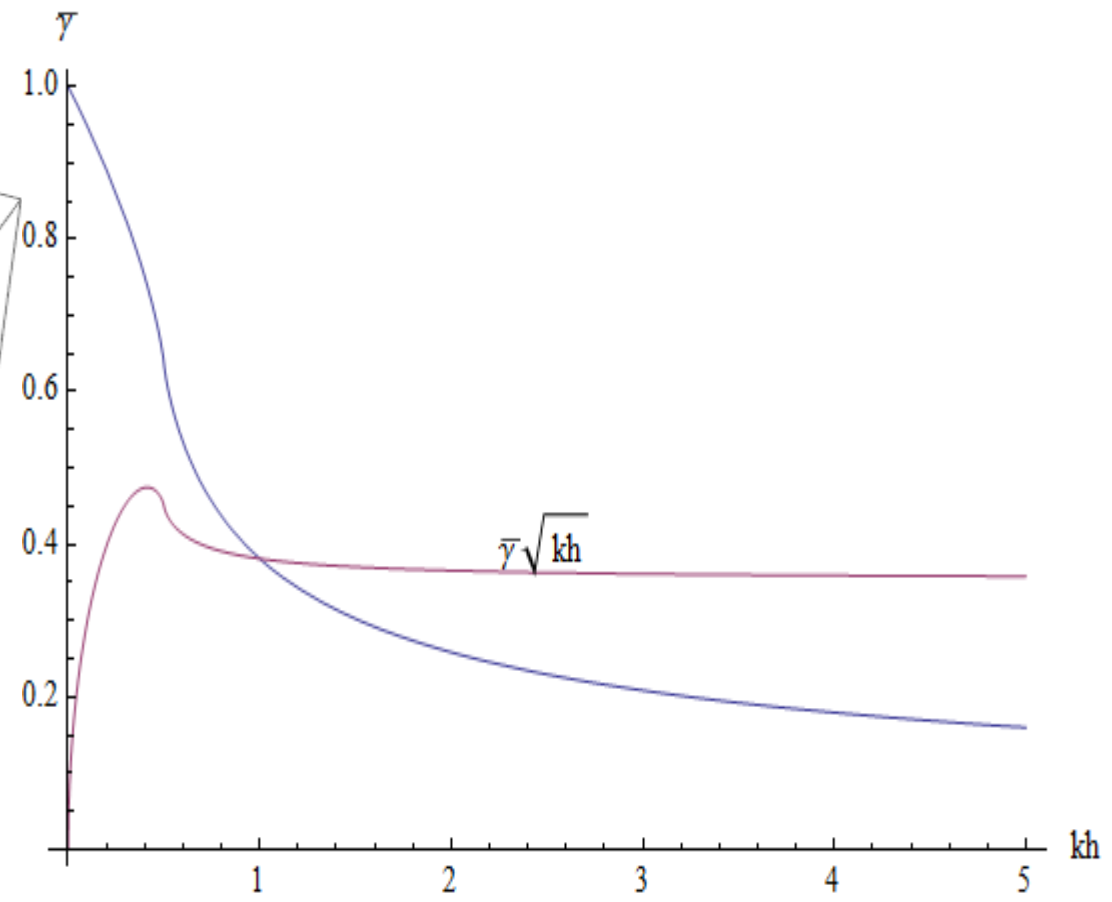

其右端极限：$kh \to \infty$ 时，$\bar{\gamma} \to \frac{\sqrt{2}}{4} / \sqrt{kh}$，即 $\bar{\Gamma} \to \frac{\sqrt{2}}{4}\sqrt{g/h}$。不变。

其左端极限：$kh \to 0$ 时，$\bar{\gamma} \to 1$。上移。

极大值出现在：$kh = 0.41305738494835587$ 时，$\bar{\Gamma} = 0.47483176886636336\sqrt{g/h}$。左移，上移。

若 $\omega_b \gg \Gamma$，则将 $\cos^2\theta$ 替换为 1/2 即得到 $\Gamma_{1,2}^2 = gk - ghk^2 = gk - \frac{1}{2}V_A^2k^2$：

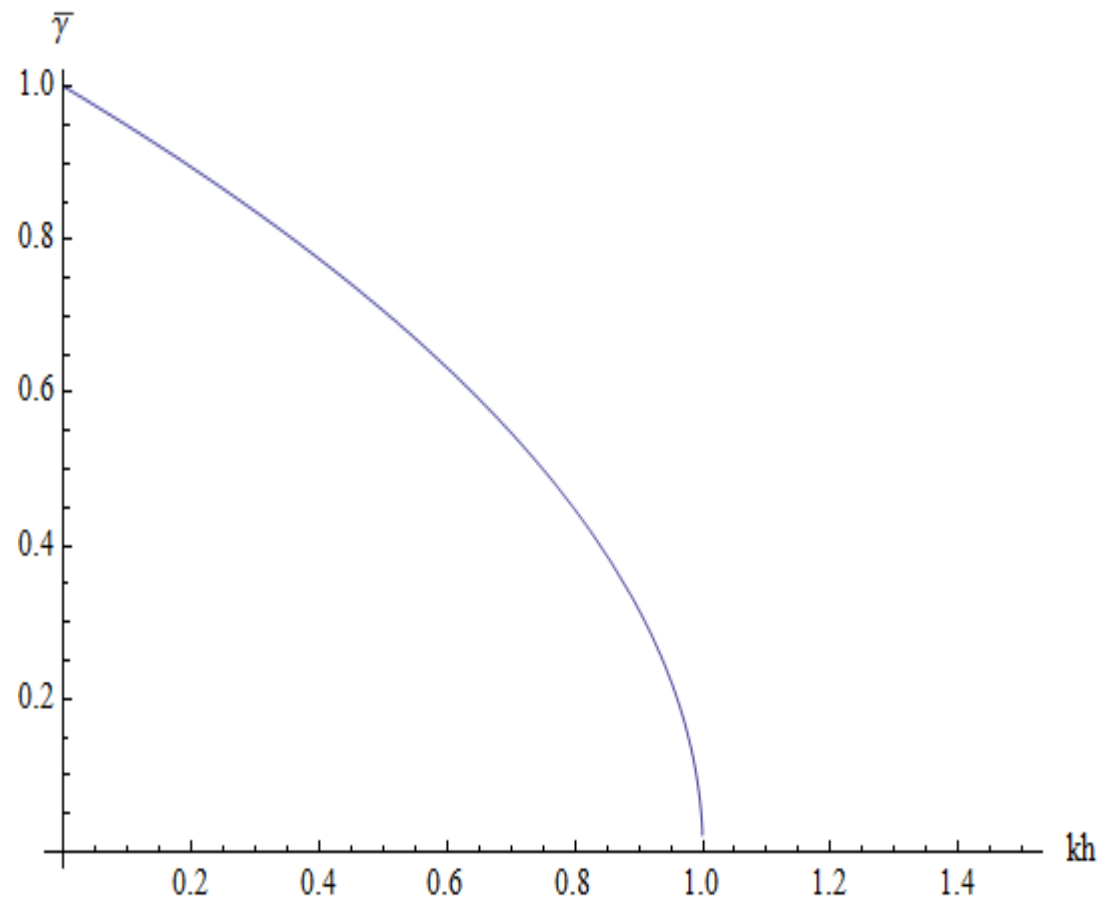

$kh \geq 1$ 的模式完全致稳。右端不变。

需要注意的是这些结果仅右端有意义，因为左端与前提条件 $kh \to \infty$ 不一致。

若退化至无限薄，$kh \to 0$，则 $R \to 2\cos^2\theta, S \to 1$，因此 $\Gamma = \sqrt{-\cos^2\theta + \sqrt{1+\cos^4\theta}}\sqrt{gk}$：

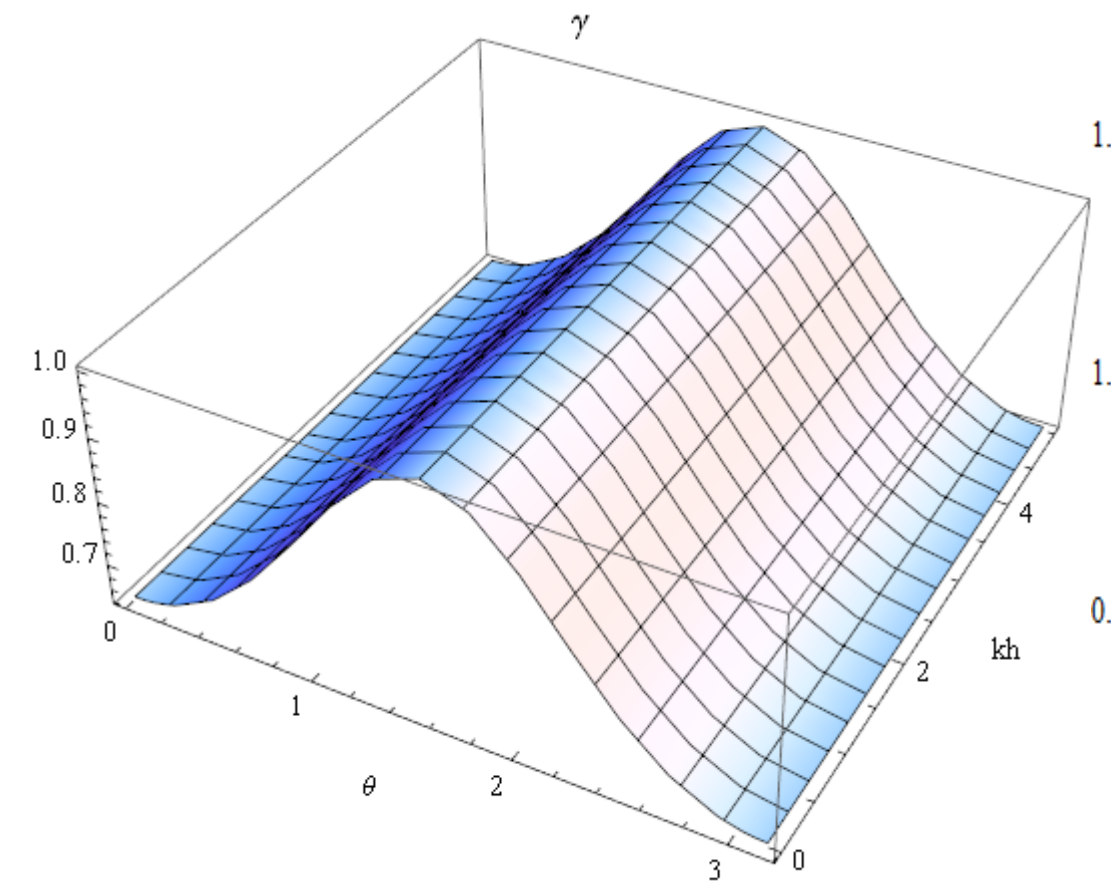

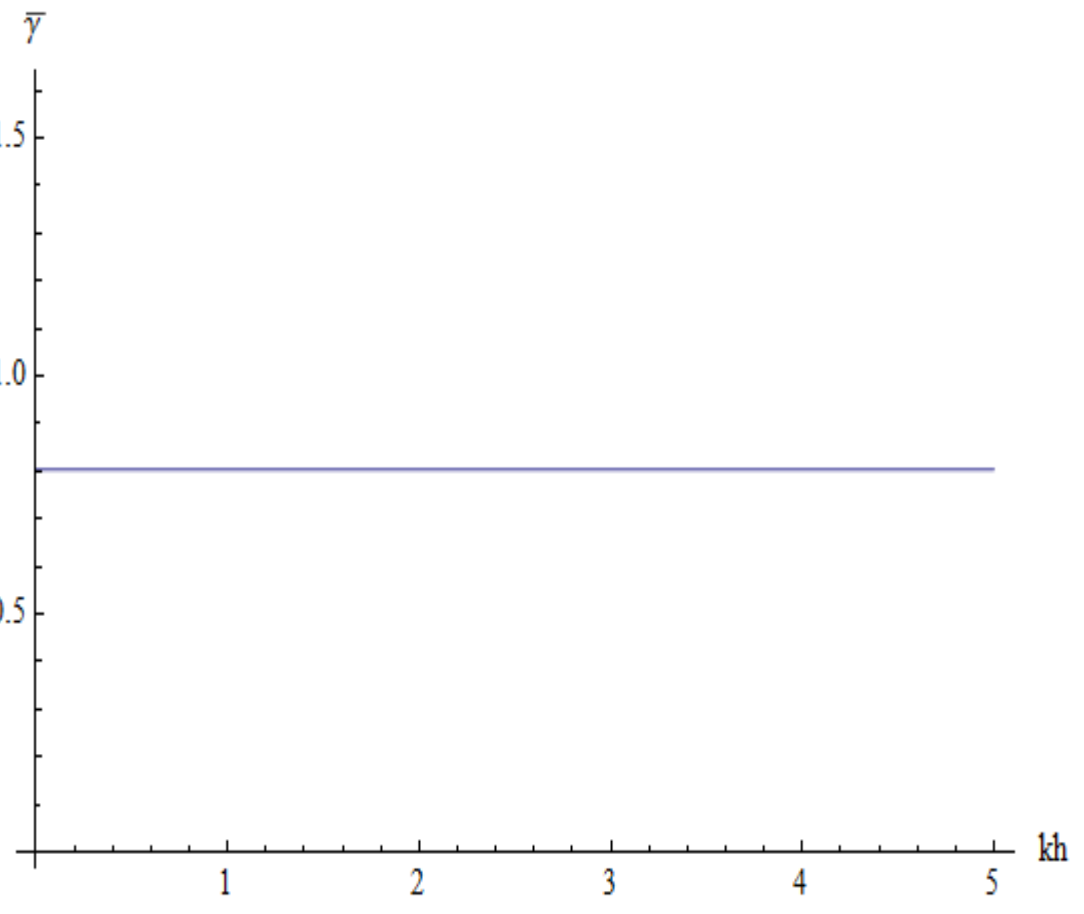

即原来结果的左端极限：$\overline{\Gamma} = 0.8041731875209983\sqrt{gk}$ 。

若 $\omega_b \gg \Gamma$，则将 $\cos^2\theta$ 替换为 1/2 即得到 $\overline{\Gamma} = \sqrt{\frac{\sqrt{5}-1}{2}}\sqrt{gk} = 0.7861513777574233\sqrt{gk}$，即原来结果的左端极限。

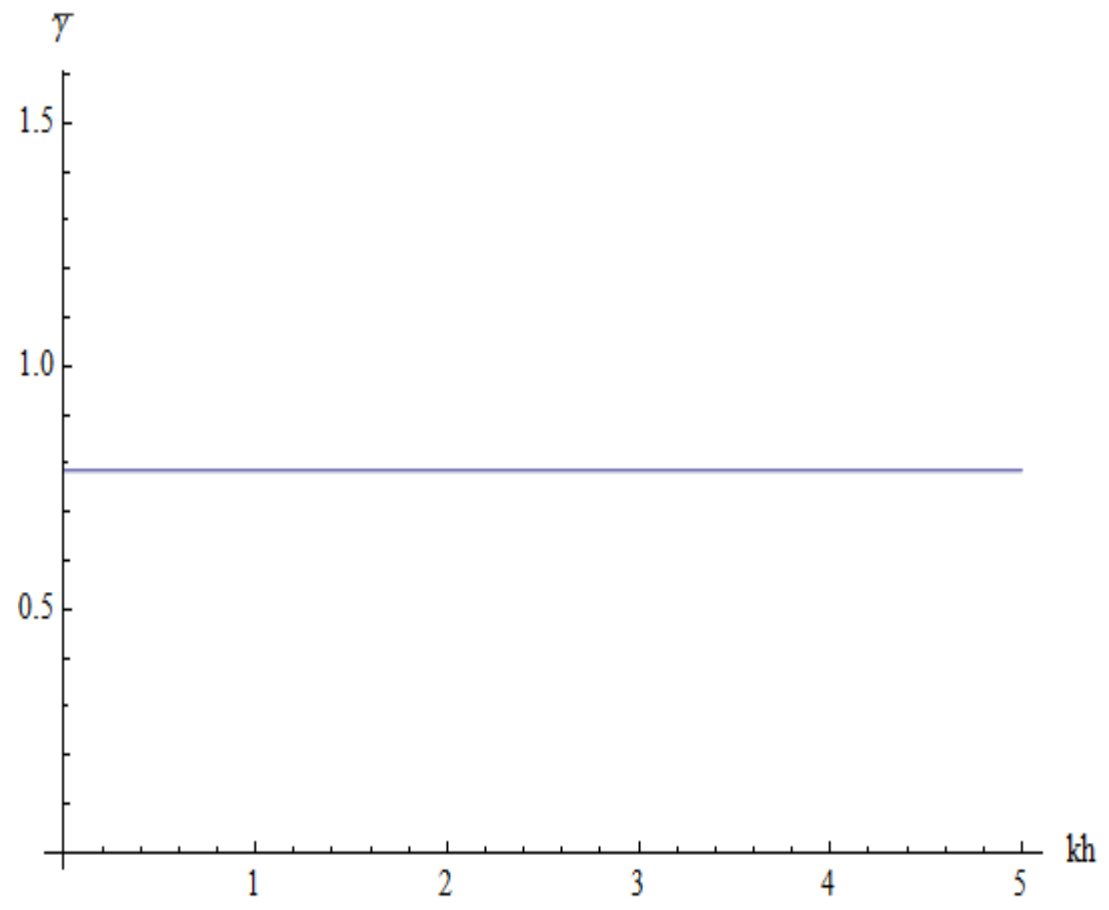

需要注意的是这些结果仅左端有意义，因为右端与前提条件 $kh \rightarrow 0$ 不一致。

**平板 non-WKB:**

取 $a \to \infty$ 极限，并假定仅驱动侧有磁场，则

$$R = 2kh\coth(kh)\cos^2\theta, S = 1 - 2kh\cos^2\theta, \tag{20}$$

其中 $\theta$ 为旋转磁场 $B(t)$ 与波矢 $k=(kx,ky,0)$ 之间夹角 $\theta(t)=\omega_b t+\theta_0$。恒有 $R \ge 0, \frac{R^2}{4} + S \ge 0$，因此 $\gamma_- = -\frac{R}{2} - \sqrt{\frac{R^2}{4} + S}$ 恒负，是振荡模式，认为/假定这个模式最终全被其他耗散机制衰减掉，舍去不予考虑，仅考虑 $\gamma_+ = -\frac{R}{2} + \sqrt{\frac{R^2}{4} + S}$ 一支：

$$(\ddot{\xi}-\gamma_{+}gk\xi)=0. \tag{21}$$

将时间用$1/\sqrt{gk}$归一化，增长率用$\sqrt{gk}$归一化，磁场方向的旋转频率用$\sqrt{g/h}$归一化，波矢用$1/h$归一化，则(20)、(21)变为：

$$R=2k\coth(k)\cos^{2}(\omega_{b}t/\sqrt{k}+\theta_{0}),S=1-2k\cos^{2}(\omega_{b}t/\sqrt{k}+\theta_{0}), \tag{22}$$

$$(\ddot{\xi}-\gamma_{+}\xi)=0. \tag{23}$$

初始条件取$\xi(0)=\xi_{0},\xi^{'}(0)=0$，数值积分(23)式至一非常大的时刻$t_{\mathrm{f}}$，例如 200。定义归一化平均增长率$\overline{\gamma}=\ln(|\xi(t_{\mathrm{f}})|/\xi_{0})/t_{\mathrm{f}}$，这样定义的增长率与$\theta_{0}$无关。$\omega_{b}=0.1,1,10$的结果为：

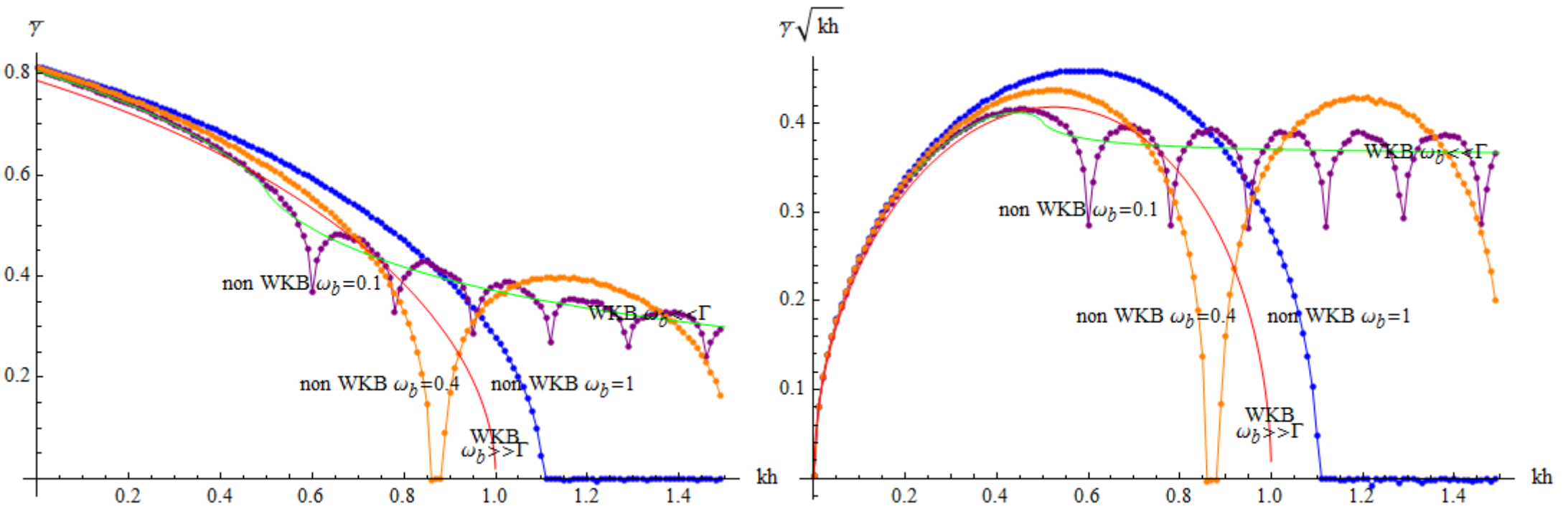

低频时退化至 WKB 低频结果。高频时，随着频率升高，第一拱门(arch)右止点右移、下移（直到 0），拱顶位置右移、上移。第一拱门右止点位置，尽管右移，似乎存在一个饱和趋近值~1.25。第一拱门拱顶位置，尽管右移、上移，似乎也存在饱和趋近值：峰位~0.63、峰值~0.47。高频时，第二拱门左起点位置近似有$k\approx\omega_{b}^{2}$，例如$\omega_{b}=2,3,5,10$的第二拱门左起点位置近似为$k\approx4,9,25,100$。第一拱门右止点与第二拱门左起点之间均是被完全抑制的。右边还有更多个高度差不多的拱门，但可以合理地假定它们能够被某种耗散机制衰减掉，耗散对高模数非常有效。

前述（归一化）平均增长率对于长时间尺度的度量是恰当的，但对于短时间尺度的度量未必适用，这时可以定义归一化最大增长率$\gamma_{\max}=\max_{\theta_{0}\in[0,2\pi],t\in[0,t_{f}]}\{\ln(|\xi(t)|/\xi_{0})/t\}$。

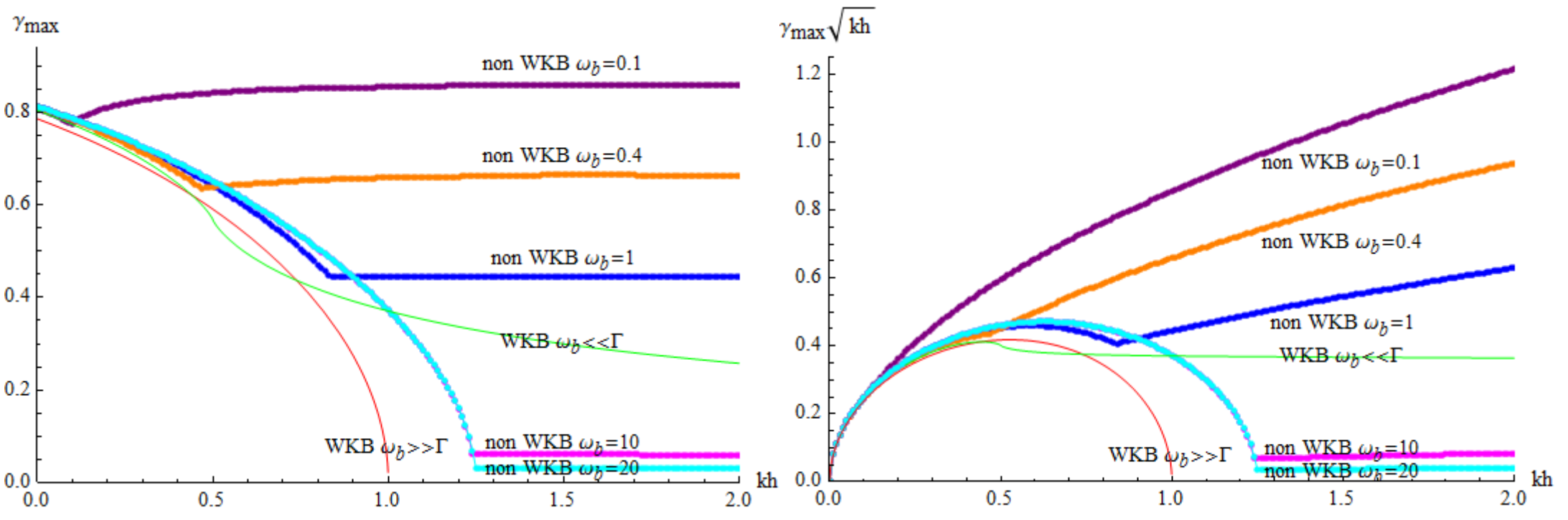

$\bar{\gamma}$ 与 $\gamma_{\max}$ 的比较，以 $\omega_b = 1$ 为例：

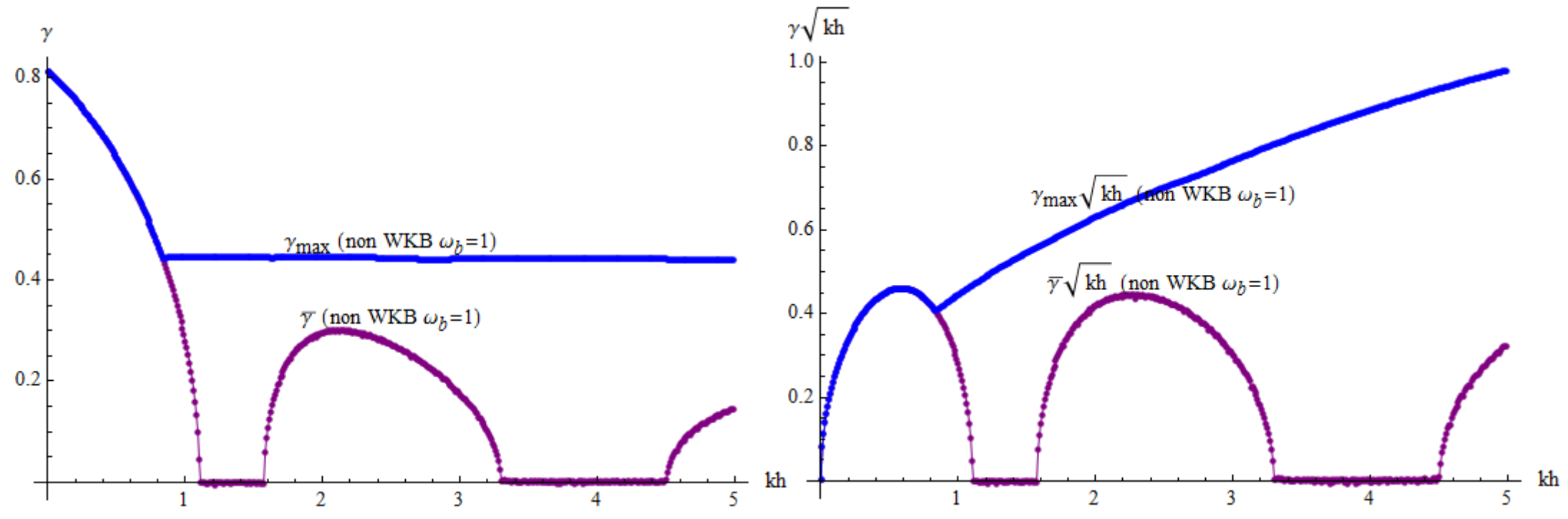

……